\documentclass[aps,prc,reprint,twocolumn,floatfix,amsmath,amssymb,superscriptaddress]{revtex4-2}
\usepackage{graphicx}
\usepackage{booktabs}
\usepackage{dcolumn}
\usepackage{microtype}
\usepackage[colorlinks=true,linkcolor=blue,citecolor=blue,urlcolor=blue]{hyperref}

\ifdefined\pdfsuppressptexinfo\pdfsuppressptexinfo=-1\fi

\begin{document}
\newcommand{\capSn}{Golden-rule coupling of the $2p_{1/2}$ hole-type resonance of $^{120}$Sn (SLy4) for the three pairing shapes at the
same mean gap (cutoff scheme). (a) Neutron pairing field $\Delta_n(r)$ (HFBRAD sign convention). (b) Integrand $I(r)$ of
Eq.~(\ref{eq:GR}). (c) Running integral, whose value at large $r$ is $M=-0.277$ (volume), $1.207$ (surface), and $0.060$ (mixed)
MeV\,fm$^{1/2}$. The dotted line marks the sign change of $I(r)$ at $5.17$~fm, which separates the interior and surface lobes.}

\newcommand{\capReg}{Cutoff versus regularized pairing for the $^{120}$Sn $2p_{1/2}$ hole (SLy4). (a) Pole widths of the three shapes
versus the cutoff energy at fixed $\langle\Delta_n\rangle=1.2448$~MeV: regularized (filled circles, solid) and cutoff scheme (open
triangles, dashed). The regularized points extend to $E_{\max}=160$~MeV. Open circles at $40$~MeV: regularized points with the hole-side term still active. (b) $\Gamma_{\rm surf}/\Gamma_{\rm vol}$: regularized (circles), cutoff at fixed $\langle\Delta_n\rangle$ (triangles) and at fixed $\Delta^{(3)}$ (squares). At
$E_{\max}=60$~MeV the box spreads (boxes of $25$, $30$, and $35$~fm) of the regularized mixed width in (a) and of the regularized
ratio in (b) are $5.2\%$ and $2.7\%$. (c) Position $\eta^*$ of the
zero of $M$ in the cutoff and regularized $\eta$ families. The bars indicate the interpolation uncertainty, and the dotted line marks
mixed pairing.}

\newcommand{\capZero}{Families of pairing shapes $\eta$ [Eq.~(\ref{eq:pair})] at fixed $\langle\Delta_n\rangle=1.2448$~MeV for the
$^{120}$Sn $2p_{1/2}$ hole: cutoff scheme at $E_{\rm cut}=40$, $60$, and $80$~MeV (open symbols, dashed) and regularized scheme at $E_{\max}=60$ and $100$~MeV (solid lines, with large open circles at $60$~MeV in (a) and (b) and filled squares at $100$~MeV, the two
nearly coinciding). (a) Pole width. Near the zero the sampled widths are upper bounds on the minimum. (b) Golden-rule matrix element
$M$. (c) $M$ versus the surface-to-core ratio $Q_2$ of $\Delta_n(r)$, where the three cutoff families coincide near the zero. The
dotted lines mark mixed pairing.}

\newcommand{\capChan}{Largest-to-smallest ratio of the pole widths among volume, surface, and mixed pairing versus the smallest
cancellation index $C_{\min}$ for all $28$ hole-type poles with $E_r>0$ in the search window of $^{120,{}116,{}124}$Sn (SLy4) and $^{70}$Zn (SkM*),
cutoff scheme. Open symbols mark channels where first-order perturbation theory deviates by more than $10\%$. The dashed line
marks the ratio $100$, and the dotted line at $C_{\min}=0.15$ is a guide to the eye. All channels with a ratio above $100$ lie to its left. The labels give the hole orbitals discussed in the text.}

\newcommand{\capPei}{Widths of hole-type resonances: Jost pole width (black), consistent golden rule (green) and golden rule with a constant
$\hbar^2/2m=20.734$~MeV\,fm$^2$ in the continuum function, as in Ref.~\cite{PKN11} (red). Left: $^{120}$Sn (SLy4) $2p_{1/2}$ for the
three shapes at the calibration strengths. Right: $^{70}$Zn (SkM*, mixed pairing) holes $1s_{1/2}$, $1p_{3/2}$, $1p_{1/2}$, $1d_{5/2}$. Open symbols: Table~I of Ref.~\cite{PKN11} (box stabilization, smoothing--fitting (smf), and golden rule). The consistent golden rule agrees with the pole width within $2.6\%$ in all seven cases.}

\title{Pairing-functional dependence of the escape widths\\
of deep-hole quasiparticle resonances in the Jost-function\\
Hartree--Fock--Bogoliubov approach with self-consistent Skyrme fields}

\author{T. Dieu Thuy}
\affiliation{Faculty of Physics, University of Education, Hue University, 34 Le Loi Street, Hue City, Vietnam}
\affiliation{Center for Theoretical and Computational Physics, University of Education, Hue University, 34 Le Loi Street, Hue City, Vietnam}
\affiliation{Phu Xuan University, 28 Nguyen Tri Phuong Street, Hue City, Vietnam}
\author{T. V. Nhan Hao}
\email[Corresponding author: ]{tvnhao@hueuni.edu.vn}
\affiliation{Faculty of Physics, University of Education, Hue University, 34 Le Loi Street, Hue City, Vietnam}
\affiliation{Center for Theoretical and Computational Physics, University of Education, Hue University, 34 Le Loi Street, Hue City, Vietnam}

\begin{abstract}
Deep neutron holes in open-shell nuclei acquire an escape width because the pairing field couples them to the particle continuum.
The escape widths of such holes in $^{116,{}120,{}124}$Sn and $^{70}$Zn are calculated from the poles of the Jost function of the
Hartree--Fock--Bogoliubov equations on self-consistent Skyrme fields. Their dependence on the radial shape of the pairing field and on
the treatment of the pairing space is investigated at fixed mean gap. With an energy cutoff and a strength renormalized to the mean
gap, the surface-pairing width of the $^{120}$Sn $2p_{1/2}$ hole grows by $31\%$ when the cutoff is doubled, and the mixed-pairing
width by a factor of $6.4$. With the regularization of Bulgac and Yu, no cutoff dependence of the volume and mixed widths of this hole is
resolved within their box spreads. The well-conditioned surface-pairing widths converge more
slowly and vary by at most $1.5\%$ between cutoffs of $120$ and $160$~MeV. The
strong sensitivity is traced to a zero of the golden-rule matrix element as a function of the radial shape of the pairing field, near
which the width drops by orders of magnitude. A cancellation index is introduced to identify such ill-conditioned widths. A golden rule
in which the continuum function is generated by the same Hamiltonian as the hole reproduces the pole widths of four deep $^{70}$Zn
holes within $0.3\%$. With a constant $\hbar^2/2m$ in the continuum function, as in the study of Pei, Kruppa, and Nazarewicz, the
perturbative widths of the two $1p$ holes are too small by more than an order of magnitude, whereas that of the $1s_{1/2}$ hole is only $14\%$ too large.

\end{abstract}

\maketitle

\section{Introduction}
\label{sec:intro}
In a superfluid nucleus described by Hartree--Fock--Bogoliubov (HFB) theory, a neutron orbital that is bound in the
Hartree--Fock (HF) field but lies deep below the Fermi energy $\lambda$
is not a stationary state. The pairing field $\Delta(r)$ couples its hole component to the particle continuum at the same
quasiparticle energy, and the state becomes a quasiparticle resonance with a finite width~\cite{DFT84,DNW96}.
Following Ref.~\cite{MQTH21}, we call such poles hole-type. Poles that grow out of bound particle states are particle-type,
and poles of the HF scattering matrix itself are shape resonances. To first order in $\Delta$ the width is a golden-rule expression
involving the matrix element of $\Delta(r)$ between the hole state and a continuum state~\cite{KM16,PKN11}. For a given radial
shape of $\Delta(r)$ it scales with the square of the average gap~\cite{KM16,ZQ20}, and Kobayashi and Matsuo have mapped out where this
perturbative description holds, down to shallow holes near the drip line~\cite{KM16,KM20}.

For deep holes this width due to pairing, which we call the escape width $\Gamma_{\rm esc}$ (the part of the decay that proceeds by
direct emission of the neutron~\cite{GSV88}), is only a small part of the observed width. As quoted in Ref.~\cite{KM16}, typical
estimates are $1$--$100$~keV~\cite{BSTF87}, ``much smaller than the experimentally known typical width (several MeV) of the deep-hole
resonances'', which is dominated by the spreading into more complex configurations. However, the radial shape of $\Delta(r)$, and not
only its average, enters the amplitude linearly. Therefore, $\Gamma_{\rm esc}$ is far more sensitive to the pairing functional than
energies and gaps, and its calculation is a stringent test of the treatment of the pairing space.

The dependence of the width on the pairing force was recognized early. Dobaczewski \textit{et al.}~\cite{DNW96} listed the
widths of deep-hole states among the observables that reflect the spatial character of pairing. Following Ref.~\cite{BSTF87}, it was noted
that these widths are in general larger for surface than for volume pairing but depend strongly on the angular momentum and excitation
energy of the state, and estimates ``based on self-consistent HFB solutions containing pairing
correlations'' were called for~\cite{DNW96}. Pei, Kruppa, and Nazarewicz found that perturbative widths of deep $p$ holes in $^{70}$Zn deviate
strongly from box-stabilized ones and traced the deviation to the surface-peaked neutron pairing potential, for which the overlap
integral ``is too sensitive to small changes'' of the continuum function~\cite{PKN11}. A second question concerns the
pairing space. A contact force requires a cutoff, and strength and cutoff together define the interaction~\cite{BDNS06}. Bulgac and Yu
replaced the cutoff by a regularization that removes the divergence of the pairing density~\cite{BY02}, and Borycki
\textit{et al.}\ compared it with renormalization for energies, gaps, and radii~\cite{BDNS06}. However, to our knowledge, the effect of
the pairing scheme on escape widths has not been examined.

Quasiparticle resonances have been studied in numerous continuum HFB calculations in coordinate space~\cite{FTZ00,Grasso01,Michel08,OM09,ZMM11,ZMM12,Sun19}.
On self-consistent Skyrme fields the Green's-function method has been used to separate the roles of the HF and pairing potentials in
neutron-rich Ca isotopes~\cite{QZ19,ZQ20}, and BCS or Bogoliubov models with Woods--Saxon (WS) fields have been studied
extensively~\cite{HM03,HM04,ISV06,KM16,KM20,KM23}. The Jost-function formulation of the HFB equations~\cite{MLTH19} gives the $S$ matrix
and its poles directly. With WS fields it has been used to classify quasiparticle resonances~\cite{MQTH21}, combined with complex
scaling~\cite{Miz26PRC}, and applied to the Fano profiles that hole-type resonances produce in elastic neutron
scattering~\cite{MLH20}. In all these works $\Delta(r)$ had a single WS form. The method has also been extended to
complex potentials~\cite{MTT23} and to the random-phase approximation (RPA), where the poles give the complex eigenenergies
of the giant dipole resonance of $^{16}$O~\cite{MLTTTH23}, and the strength function has been decomposed into eigenphase-shift
components~\cite{MTH24} and into pole contributions~\cite{MT25}. The extensions to the HFB and RPA frameworks are reviewed in Ref.~\cite{Miz26FBS}.

The Jost-function approach has several features that make it well suited to the present problem. First, the outgoing
boundary condition is imposed exactly in both quasiparticle channels, so that a resonance is a zero of $\det\mathcal J^{(+)}(E)$ on a
definite Riemann sheet, and its energy and width follow without any fit or smoothing of a discretized spectrum. Second, the pole search does not rely on resolving the resonance on an energy grid, so that very narrow and broad resonances are treated in the same way. Third, the same Jost
matrices give the $S$ matrix on the real energy axis, which provides an independent determination of the resonance parameters and
connects the poles with observables such as the Fano profiles of elastic neutron scattering~\cite{MLH20}. Finally, a pole can be
followed continuously as the pairing field is switched on, which identifies it as hole-type or particle-type~\cite{MLTH19,MQTH21} and gives
the perturbative limit of its width directly. Methods based on box discretization obtain widths from the box-size dependence or from
smoothed occupation numbers of discretized states, and their application to very narrow resonances is difficult~\cite{PKN11}.

The goal of the present work is to compute hole-type poles on fully self-consistent Skyrme--HFB fields and to determine how
$\Gamma_{\rm esc}$ depends on the pairing functional, namely on the radial shape of $\Delta(r)$ and on the treatment of the pairing
space. We compare a cutoff scheme with a renormalized strength and the regularization of Bulgac and Yu, and we
follow the poles along a continuous family of pairing shapes at fixed mean gap. The strong sensitivities found in this way are traced
to a zero of the golden-rule matrix element, and a cancellation index, which quantifies the sensitivity noted in Ref.~\cite{PKN11},
signals the proximity of such a zero. Finally, we revisit the perturbative widths of Ref.~\cite{PKN11} and identify the
ingredient of the golden rule that is responsible for their discrepancy with the box-stabilized widths.

\section{Formalism}
\label{sec:form}

\subsection{Self-consistent fields and the two pairing schemes}
\label{sec:form-fields}
The fields of the even--even cores are obtained with the coordinate-space HFB code HFBRAD~\cite{BD05} in spherical symmetry, with
SLy4~\cite{SLy98} for the tin isotopes and SkM*~\cite{SkMs} for $^{70}$Zn. The pairing interaction is the zero-range,
density-dependent force of HFBRAD~\cite{BD05},
\begin{equation}
V_{\rm pair}(\mathbf r_1,\mathbf r_2)=t_0'\Big[1-\eta\,\frac{\rho(\mathbf r_1)}{\rho_0}\Big]\delta(\mathbf r_1-\mathbf r_2)
\equiv G(\mathbf r_1)\,\delta(\mathbf r_1-\mathbf r_2),
\label{eq:pair}
\end{equation}
where $\rho$ is the isoscalar density and $\rho_0=0.16$~fm$^{-3}$, i.e., $t_3'=-6\eta t_0'/\rho_0$ with the power $\gamma'=1$. Volume, surface, and mixed pairing correspond to $\eta=0$, $1$, and $0.5$. Intermediate values of $\eta$ define a continuous family of shapes, which is used in
Sec.~\ref{sec:res-zero}.

\emph{Cutoff scheme.} The neutron pairing field is $\Delta_n(r)=\tfrac12 G(r)\,\tilde\rho_n(r)$, where the pairing density
$\tilde\rho_n$ is summed over quasiparticle states $k$ with a smooth Fermi-type cutoff of diffuseness $d$. As in HFBRAD, the cutoff acts
on the equivalent single-particle energy $\bar e_k=\lambda+(1-2v_k^2)E_k$ of Ref.~\cite{BDNS06}, $v_k^2$ being the norm of
the lower component, so that $E_{\rm cut}$ is an absolute energy on the scale of $\lambda$. A contact force requires a limited pairing
space and a renormalized strength~\cite{DNW96}. We refit the strength for each cutoff to the same mean gap, which is the
renormalization of Refs.~\cite{DNW96,BDNS06}.

\emph{Regularized scheme.} Following Bulgac and Yu~\cite{BY02}, HFBRAD~\cite{BD05} replaces $G$ by an effective
coupling,
\begin{equation}
\begin{split}
&\Delta_n(r)=\tfrac12\,g_{\rm eff}(r)\,\tilde\rho_n(r),\qquad \frac1{g_{\rm eff}(r)}=\frac1{G(r)}-\alpha(r),\\
&\alpha(r)=\frac{m^*(r)\,k_c(r)}{2\pi^2\hbar^2}\Big[1-\frac{k_F(r)}{2k_c(r)}\ln\frac{k_c(r)+k_F(r)}{k_c(r)-k_F(r)}\Big],
\end{split}
\label{eq:reg}
\end{equation}
with $\hbar^2k_F^2/2m^*+U=\lambda$ and $\hbar^2k_c^2/2m^*+U=E_{\max}$ defined locally, the position-dependent mass of
Ref.~\cite{BDNS06}, and a sharp cutoff of the sum at $\bar e_k=E_{\max}$. Where $U(r)>\lambda$, $k_F=i\kappa_F$ is imaginary and the bracket in Eq.~(\ref{eq:reg}) is continued analytically to $1+(\kappa_F/k_c)\arctan(\kappa_F/k_c)$, and where $U(r)>E_{\max}$ the coupling is not modified. In the notation of Ref.~\cite{BY02} this
is the quasiparticle cutoff $E_c=E_{\max}-\lambda_n$, and $G$ plays the role of their coupling $g$. Their pairing density refers to one
spin component, whereas $\tilde\rho_n$ sums both, hence the factor $\tfrac12$ in Eq.~(\ref{eq:reg}). In the conventions of
Refs.~\cite{BD05} and \cite{BDNS06} the same coupling reads $G/4$ and $G/2$. The hole-side term of
Ref.~\cite{Bul02} is also included, and it vanishes once $E_{\max}$ exceeds $2\lambda_n-\min U_n$, about $57$~MeV here. Since the divergent part
of the pairing density is removed, observables should become independent of $E_{\max}$ once it is large~\cite{BY02,BDNS06}. At the same
$\eta$ the two schemes do not give the same radial shape of $\Delta_n(r)$, because $g_{\rm eff}/G$ depends on $r$.

The density dependence of $G$ contributes a rearrangement term to the central field of both nucleon species. In the
regularized scheme the pairing energy is $E_{\rm pair}=\tfrac14\sum_q\int g_{{\rm eff},q}\,\tilde\rho_q^{\,2}\,d^3r$, and its functional
derivative with respect to the density at fixed $\alpha$ gives
\begin{equation}
U_{\rm rearr}(r)=\frac14\,\frac{\partial G}{\partial\rho}\sum_{q=n,p}\frac{\tilde\rho_q^{\,2}(r)}{[1-\alpha_q(r)\,G(r)]^2},
\label{eq:rearr}
\end{equation}
with $\partial G/\partial\rho=-\eta\,t_0'/\rho_0$, where $\alpha_q$ is the function $\alpha$ of Eq.~(\ref{eq:reg}) for nucleon species $q$
and $(1-\alpha_qG)^{-2}=(g_{{\rm eff},q}/G)^2$. This form is also regular where $G$ vanishes. This corresponds to the first term of Eq.~(2.12) of Ref.~\cite{BDNS06}. The terms
arising from the density dependence of $\alpha$, through $k_F$, $k_c$, and the factor $m^*$ in Eq.~(\ref{eq:reg}), are neglected. Those
through $k_F$ and $k_c$ are also neglected in Ref.~\cite{BD05}, and for volume and mixed pairing they were
found to change the total energy by a few keV~\cite{BDNS06}. Their size for surface pairing is not estimated in the present work. The factor $(1-\alpha_qG)^{-2}$ is not present in the version of
HFBRAD distributed with Ref.~\cite{BD05} and has been added in the present calculations. In the cutoff scheme $\alpha_q=0$, and the factor
is unity. For volume pairing $\partial G/\partial\rho=0$, and the term vanishes in both schemes.

\emph{Anchors.} The mean gap is the pairing field averaged with the particle density,
\begin{equation}
\langle\Delta_n\rangle=-\frac{\int\Delta_n(r)\,\rho_n(r)\,r^2dr}{\int\rho_n(r)\,r^2dr},
\label{eq:gap}
\end{equation}
normalized with the neutron number as in Refs.~\cite{DNW96,BDNS06}. The minus sign accounts for the sign convention of
HFBRAD, so that $\langle\Delta_n\rangle>0$. Alternatively, the strength
is fixed by the three-point odd--even mass difference~\cite{SDN98},
\begin{equation}
\Delta^{(3)}(N)=\tfrac12\big[2E(N)-E(N-1)-E(N+1)\big],
\label{eq:d3}
\end{equation}
computed from HFB total energies of $^{118,{}120}$Sn and of $^{119}$Sn blocked in the equal-filling approximation~\cite{PMR08}, which we
implemented in HFBRAD. The experimental value is
$\Delta^{(3)}(69)=1.3103(9)$~MeV~\cite{AME20}. Blocking is not combined with regularization in the present calculations, so the regularized
scheme is anchored to $\langle\Delta_n\rangle$ only. The two anchors are not equivalent~\cite{BRRM00b}.

\subsection{Jost function and hole-type poles}
\label{sec:form-jost}
For a given core field and a constant nucleon mass, the radial HFB equation for the quasiparticle wave function
$\phi_{lj}=(\varphi_1,\varphi_2)^{T}$ reads~\cite{MLTH19}
\begin{equation}
\Big[\Big(\frac{d^2}{dr^2}-\frac{l(l+1)}{r^2}\Big)\mathbf 1-\hat S_z\frac{2m}{\hbar^2}\,\mathcal U_{lj}(r)\Big]\phi_{lj}
=-\begin{pmatrix}k_1^2&0\\0&k_2^2\end{pmatrix}\phi_{lj},
\label{eq:HFB}
\end{equation}
with $\mathcal U_{lj}=\begin{pmatrix}U_{lj}&\Delta\\ \Delta&-U_{lj}\end{pmatrix}$, $\hat S_z={\rm diag}(1,-1)$,
$k_1=\sqrt{2m(\lambda+E)}/\hbar$, and $k_2=\sqrt{2m(\lambda-E)}/\hbar$~\cite{MLTH19}. The mean field and $\Delta(r)$ are
those of the self-consistent core. With the position-dependent mass function $B(r)=\hbar^2/2m^*(r)$ of the Skyrme functional
(denoted $M_q$ in Ref.~\cite{BD05}) the equation takes
the form of Eq.~(33) of Ref.~\cite{BD05}, and HFBRAD removes the first derivatives by an $r$-dependent transformation of the two
components~\cite{BD05}. For the contact force of Eq.~(\ref{eq:pair}) the pairing field has no effective-mass part, and the
transformation reduces to the scaling $\varphi_i=B^{-1/2}f_i$, which gives
\begin{equation}
\begin{split}
-B\,f_1''+V_{lj}\,f_1+\Delta\,f_2&=E\,f_1,\\
B\,f_2''-V_{lj}\,f_2+\Delta\,f_1&=E\,f_2,
\end{split}
\label{eq:liou}
\end{equation}
\begin{equation}
V_{lj}=U+U^{\rm so}_{lj}-\lambda+B\frac{l(l+1)}{r^2}+\frac{B'}{r}+\frac{B''}{2}-\frac{B'^2}{4B}.
\label{eq:Vlj}
\end{equation}
Here $U$ is the central field and $U^{\rm so}_{lj}$ the spin--orbit field of Ref.~\cite{BD05}, and the last two terms
of Eq.~(\ref{eq:Vlj}) come from the scaling. Equations~(\ref{eq:liou}) are integrated with the Numerov method of HFBRAD, extended to
complex energies. The fields are cut at $R_m=20$~fm, where they are negligible. Moving the cut to $18$~fm changes the $2p_{1/2}$ widths of $^{120}$Sn by less
than $2\times10^{-7}$ in relative value. Beyond $R_m$, $B$ equals its asymptotic value
$B_\infty$, and Eqs.~(\ref{eq:liou}) reduce to Eq.~(\ref{eq:HFB}) with $U_{lj}=\Delta=0$ and $\hbar^2/2m$ replaced by $B_\infty$, which
contains the factor $1-1/A$ of the center-of-mass correction of SLy4 and SkM*, so that the asymptotic momenta are
$k_1=[(\lambda+E)/B_\infty]^{1/2}$ and $k_2=[(\lambda-E)/B_\infty]^{1/2}$. The constant factor $B_\infty^{1/2}$ between $f_i$ and
$\varphi_i$ does not affect the $S$ matrix. Asymptotically, each of the two regular solutions is a combination of
incoming and outgoing waves in the two components, and the coefficients of the incoming and outgoing waves define the $2\times2$ Jost
matrices $\mathcal J^{(+)}(E)$ and $\mathcal J^{(-)}(E)$, respectively, as in Eqs.~(33) and (38) of Ref.~\cite{MLTH19}. Following
Eq.~(43) of Ref.~\cite{MLTH19}, the $S$ matrix is $S(E)=[\mathcal J^{(+)}(E)]^{-1}\mathcal J^{(-)}(E)$,
and its poles are the zeros of $\det\mathcal J^{(+)}(E)$ on the sheet ${\rm Im}\,k_1<0$, ${\rm Im}\,k_2>0$~\cite{MLTH19},
with
\begin{equation}
E_r=\lambda+{\rm Re}\,E,\qquad \Gamma=-2\,{\rm Im}\,E .
\label{eq:ErG}
\end{equation}
The sign convention for $\Gamma$ follows Fig.~1 of Ref.~\cite{MLTH19} and
Ref.~\cite{Miz26PRC}. Each hole-type pole is followed continuously from the HF hole level $e_h$ of the same field, at the
quasiparticle energy $E_h=\lambda-e_h$, as the
pairing field is switched on, scaling $\Delta(r)$ from zero to its self-consistent value. At small scale factors the width grows
quadratically. Orbitals are labeled in the usual way, $n=1$ denoting the lowest level of each partial wave. Since the pole search involves no box, widths spanning more than six orders of magnitude (Sec.~\ref{sec:res-reg}) are obtained with
the same procedure. The box dependence reported in Sec.~\ref{sec:setup} therefore comes from the self-consistent fields, which are
computed in a box, and not from the determination of the poles.

\subsection{Golden rule and cancellation index}
\label{sec:form-gr}
To first order in $\Delta$ the width of a hole-type pole is given by the golden rule~\cite{DNW96,PKN11,KM16,MLH20}, in Fano's form~\cite{Fano61},
\begin{equation}
\begin{split}
\Gamma_{\rm GR}&=2\pi\,\big|\langle\psi_{E}|\Delta|\varphi_h\rangle\big|^2=\frac{2M^2}{B_\infty\,k},\\
M&=\int_0^\infty I(r)\,dr,\qquad I(r)=\frac{u(r)}{A}\,\Delta_n(r)\,w_h(r),
\end{split}
\label{eq:GR}
\end{equation}
where $w_h$ is the HF hole wave function normalized to unity, $u$ the regular HF solution of the particle channel at the same energy,
$A$ its asymptotic amplitude, $k=[(\lambda+E_h)/B_\infty]^{1/2}$, and $\psi_E=(u/A)/(\pi B_\infty k)^{1/2}$ is normalized in energy.
Both $u$ and $w_h$ are eigenfunctions of the same HF Hamiltonian, including the position-dependent effective mass. We refer to this
as the consistent golden rule. Since $u$ and $w_h$ belong to the same partial wave and to different energies, $\lambda+E_h$ and
$\lambda-E_h$, they are orthogonal, $\int_0^\infty u\,w_h\,dr=0$. For the $^{120}$Sn $2p_{1/2}$ hole in the three cutoff fields at $E_{\rm cut}=60$~MeV used in
Sec.~\ref{sec:res-zero}, the computed integral is below $0.08\%$ of $\int_0^\infty|u\,w_h|\,dr$. A constant pairing field therefore gives $M=0$,
and $M=\int_0^\infty(u/A)\,(\Delta_n-\bar\Delta)\,w_h\,dr$ for any constant $\bar\Delta$, so that only the radial variation of
$\Delta_n(r)$ contributes to $M$. A continuum function generated by a different Hamiltonian is in general not orthogonal to $w_h$, and
a constant part of $\Delta_n(r)$ then also contributes to $M$ (Sec.~\ref{sec:res-pei}). Because $I(r)$ changes sign, the degree of cancellation in $M$ is measured by
\begin{equation}
C=\frac{|M|}{\int_0^\infty|I(r)|\,dr},\qquad 0\le C\le1 .
\label{eq:C}
\end{equation}
Since $\Gamma_{\rm GR}\propto M^2$, a relative change $\delta$ of one of two competing lobes of $I(r)$ changes the width by a relative
amount $(1\pm C)\,\delta/C$ to first order in $\delta$ (upper sign for the larger lobe), i.e., by about $\delta/C$ for small $C$, so that $1/C$ plays the role of a condition number of the width.
$M$, $C$, and the zeros of $M$ are evaluated from the HFBRAD fields by integrating the untransformed radial
equations, independently of the pole search. All widths reported below are pole widths, Eq.~(\ref{eq:ErG}), and Eq.~(\ref{eq:GR}) is
used only for their interpretation.

\subsection{Real-axis width}
\label{sec:form-real}
The widths are also evaluated on the real energy axis, independently of the pole search. With $T=(i/2)(S-1)$~\cite{MLTH19}
for the full HFB problem and $T^{(0)}$ for the HF field with $\Delta=0$, the resonant part $T^{(1)}$ of the amplitude
is given by the phase-shift difference $\delta^{(1)}=\delta-\delta^{(0)}$ in the elastic channel,
$T^{(1)}=-e^{i\delta^{(1)}}\sin\delta^{(1)}$, so that $T=T^{(0)}+T^{(1)}S^{(0)}$, and
defines~\cite{MLH20}
\begin{equation}
\epsilon=\frac{1-iT^{(1)}}{T^{(1)}} ,
\label{eq:eps}
\end{equation}
and the real-axis resonance energy and width follow from $\epsilon(E_r)=0$ and $d\epsilon/dE|_{E_r}=2/\Gamma$~\cite{MLH20}. They are
denoted by $E_r^{\epsilon}$ and $\Gamma^{\epsilon}$.

\section{Numerical setup and accuracy}
\label{sec:setup}
The HFBRAD equations are solved in a box of $30$~fm with a mesh of $h=0.2$~fm and angular momenta up to $2j_{\max}=39$ for neutrons
and $25$ for protons. In the cutoff scheme the reference window is $E_{\rm cut}=60$~MeV with $d=1.0$~MeV. We vary $E_{\rm cut}$ ($40$
and $80$~MeV), $d$ ($0.5$ and $2.0$~MeV), and $2j_{\max}$ ($31$ and $47$) one at a time, starting from the HFBRAD calibration
$t_0'=-186.5$, $-509.6$, and $-283.3$~MeV\,fm$^3$ for volume, surface, and mixed pairing~\cite{BD05}. In the regularized scheme
the sum is cut at $E_{\max}=40$, $60$, $80$, $100$, and $120$~MeV, starting from the regularized strengths $t_0'=-233.0$, $-914.2$, and $-370.2$~MeV\,fm$^3$ of the code. For the convergence test the $^{120}$Sn holes are also computed at $E_{\max}=140$, $150$, and $160$~MeV. At $E_{\max}=40$~MeV the hole-side term mentioned below Eq.~(\ref{eq:reg}) is still active, and the
regularized results are taken from $E_{\max}\ge60$~MeV. The self-consistent iterations are
stopped when the relative change of the total energy is below $10^{-9}$ and that of the summed gaps below $10^{-7}$.
For the surface field at $E_{\max}=150$~MeV and for the three fields at $160$~MeV the energy criterion is relaxed to
$10^{-7}$, since the total energy fluctuates at this level between iterations. At $E_{\max}=120$~MeV this changes the widths by less
than $1.4\times10^{-5}$.

For $^{120}$Sn the overall scale of $(t_0',t_3')$ is refitted for each shape, scheme, and window to $\langle\Delta_n\rangle=1.2448$~MeV,
the renormalization value $1.245$~MeV of Refs.~\cite{DNW96,BDNS06}. In the cutoff scheme the strength is also refitted to
$\Delta^{(3)}(69)=1.31031$~MeV. For this anchor the lowest blocked configuration of $^{119}$Sn is $d_{3/2}$ in all sets examined,
whereas the observed ground state is $1/2^+$~\cite{NDS119}. Blocking $s_{1/2}$ instead would raise $\Delta^{(3)}$ by $0.110$ to $0.134$~MeV at the $\Delta^{(3)}$-fit strengths and by $0.095$ to $0.237$~MeV at the mean-gap strengths. Therefore,
the mean gap is taken as the primary anchor. For $^{116,{}124}$Sn, and for $^{120}$Sn in Secs.~\ref{sec:res-channels} and \ref{sec:res-pei}, the calibration strengths are used unchanged, and for $^{70}$Zn (SkM*)
the volume and surface strengths are refitted to the mean gap of the calibrated mixed field, $0.9552$~MeV. No parameter is adjusted to
a resonance energy or width.

\emph{HFB fields.} With the default mixed strength our regularized $^{120}$Sn solution reproduces the total energy and the mean gap
of Ref.~\cite{BD05} within the spreads quoted there ($\pm16$ and $\pm7$~keV). The regulator is that of Eq.~(18) of Ref.~\cite{BY02}.
With the rearrangement term of Eq.~(\ref{eq:rearr}), the central field agrees with the functional derivative of the total energy at
fixed $\alpha$, obtained from small local variations of the density, within a relative deviation of $8.1\times10^{-4}$, against
$7.9\times10^{-4}$ in the cutoff scheme.

\emph{Accuracy of the widths.} The pole widths agree with the real-axis evaluation of Sec.~\ref{sec:form-real}, which does
not involve the pole search, within $0.14\%$, and the energies within $2.4$~keV, in all $84$ hole channels of the tin and zinc fields.
When the pairing field is scaled by $s=0.05$, the golden rule, Eq.~(\ref{eq:GR}), reproduces the pole widths divided by $s^2$ within
$2.0\%$ for every channel and shape of Sec.~\ref{sec:res-channels}. Since $M$ is evaluated from the untransformed radial equations
(Sec.~\ref{sec:form-gr}), this agreement also validates the transformation leading to Eqs.~(\ref{eq:liou}). At full strength, $C$ is
used only where the golden rule is within $10\%$ of the pole width for all three shapes.

\emph{Mesh and partial waves.} Halving the mesh changes $\Gamma/s^2$ at $s=0.05$ by at most $0.45\%$ in the $24$ channel--shape
combinations of $^{120}$Sn in the cutoff scheme, and the regularized $2p_{1/2}$ widths at $E_{\max}=60$~MeV by at most $1.3\%$.
Increasing $2j_{\max}$ from $39$ to $47$ changes these widths by at most $0.020\%$.

\emph{Box.} The box is the numerical parameter with the largest effect. The box spreads, i.e., the variations of a width over boxes
of $25$, $30$, and $35$~fm, of all $^{120}$Sn widths at $E_{\max}=E_{\rm cut}=60$~MeV are given in Sec.~\ref{sec:res-reg}. For the
regularized $2p_{1/2}$ widths at $E_{\max}=120$~MeV they are $1.5\%$, $0.56\%$, and $1.9\%$, and at fixed $\Delta^{(3)}$ the cutoff
box spreads reach $7.7\%$. A cutoff dependence of a width is considered as resolved only when it exceeds the box spread of the same
channel, shape, and scheme.

\section{Results}
\label{sec:res}
\subsection{Cutoff and regularized pairing}
\label{sec:res-reg}

\begin{table}[t]
\caption{Pole widths (keV) of the $^{120}$Sn $2p_{1/2}$ hole-type resonance and their ratios in the two pairing schemes. Regularized:
sharp cutoff at $E_{\max}$, strength refitted to $\langle\Delta_n\rangle=1.2448$~MeV. Cutoff: smooth cutoff $E_{\rm cut}$ with
$d=1.0$~MeV, strength refitted to $\langle\Delta_n\rangle$ or to $\Delta^{(3)}(69)=1.31031$~MeV. $E$ denotes $E_{\max}$ or
$E_{\rm cut}$. The ratios $R_{\rm sv}=\Gamma_{\rm surf}/\Gamma_{\rm vol}$ and $R_{\rm sm}=\Gamma_{\rm surf}/\Gamma_{\rm mix}$ are
computed from unrounded widths. At fixed $\Delta^{(3)}$ and $E_{\rm cut}=60$~MeV, $\langle\Delta_n\rangle=1.743$, $1.187$, and $1.585$~MeV (vol, surf, mix). The
box spread of the regularized widths is $0.45$--$5.2\%$ at $E_{\max}=60$~MeV. $^{a}$Hole-side term still active (Sec.~\ref{sec:setup}). $^{b}$Energy criterion of the iterations relaxed to $10^{-7}$ (Sec.~\ref{sec:setup}).}
\label{tab:reg}
\setlength{\tabcolsep}{3.2pt}
\begin{ruledtabular}
\begin{tabular}{lrrrrrr}
$E$ (MeV) & $\Gamma_{\rm vol}$ & $\Gamma_{\rm surf}$ & $\Gamma_{\rm mix}$ & $R_{\rm sv}$ & $R_{\rm sm}$ & $C_{\rm mix}$ \\
\hline
\multicolumn{7}{l}{regularized, fixed $\langle\Delta_n\rangle$} \\
$40^{a}$ & 27.52 & 111.8 & 3.373 & 4.063 & 33.16 & 0.1187 \\
$60$     & 27.97 & 109.6 & 3.587 & 3.920 & 30.57 & 0.1223 \\
$80$     & 27.70 & 108.4 & 3.578 & 3.912 & 30.28 & 0.1222 \\
$100$    & 28.08 & 107.9 & 3.686 & 3.841 & 29.26 & 0.1239 \\
$120$    & 27.92 & 107.0 & 3.677 & 3.834 & 29.11 & 0.1238 \\
$140$ & 28.29 & 106.9 & 3.773 & 3.779 & 28.34 & 0.1253 \\
$150^{b}$ & 27.99 & 107.2 & 3.691 & 3.829 & 29.04 & 0.1240 \\
$160^{b}$ & 28.27 & 106.5 & 3.793 & 3.768 & 28.09 & 0.1255 \\
\multicolumn{7}{l}{cutoff, fixed $\langle\Delta_n\rangle$} \\
$40$ & 16.95 & 282.7 & 0.2564 & 16.68 & 1102.8 & 0.0300 \\
$60$ & 16.45 & 327.5 & 0.8259 & 19.91 & 396.6  & 0.0541 \\
$80$ & 15.84 & 369.6 & 1.636  & 23.33 & 225.9  & 0.0759 \\
\multicolumn{7}{l}{cutoff, fixed $\Delta^{(3)}$} \\
$40$ & 32.66 & 283.3 & 0.4590 & 8.68 & 617.4 & 0.0311 \\
$60$ & 32.46 & 296.5 & 1.531  & 9.13 & 193.7 & 0.0577 \\
$80$ & 31.17 & 310.2 & 3.011  & 9.95 & 103.0 & 0.0816 \\
\end{tabular}
\end{ruledtabular}
\end{table}

\begin{figure*}[t]
\includegraphics[width=\textwidth]{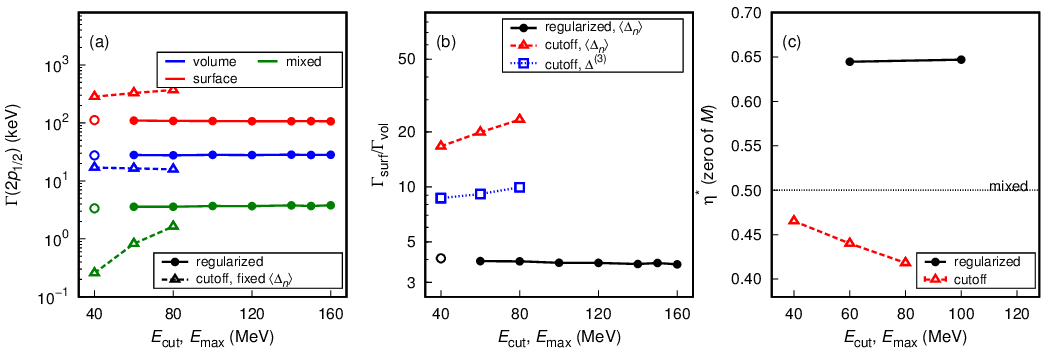}
\caption{\capReg}
\label{fig:reg}
\end{figure*}

\begin{table*}[t]
\caption{Regularized pairing for all eight hole channels of $^{120}$Sn (SLy4) at fixed $\langle\Delta_n\rangle=1.2448$~MeV: pole widths (keV) at $E_{\max}=60$ and $120$~MeV, spread $\delta_E$ ($\%$, max/min$\,-1$) over $E_{\max}=60$, $80$, $100$, and $120$~MeV, and box spreads ($\%$, boxes of $25$, $30$, and $35$~fm) of the regularized ($E_{\max}=60$~MeV) and cutoff ($E_{\rm cut}=60$~MeV) widths. $C$: cancellation index of the regularized field at $E_{\max}=60$~MeV. $^{a}$Near a zero of $M$, where the entry only shows that the cell is ill-conditioned. $^{b}$$\delta_E$ larger than the regularized box spread. The superscript $(2)$ marks the second pole of a partial wave in order of increasing $E_h$.\label{tab:regall}}
\begin{ruledtabular}
\begin{tabular}{llrrrrrr}
hole & & $\Gamma(60)$ & $\Gamma(120)$ & $\delta_E$ & $\delta^{\rm reg}_{\rm box}$ & $\delta^{\rm cut}_{\rm box}$ & $C$ \\
\hline
$2p_{1/2}$ & vol & 27.969 & 27.918 & 1.4 & 2.2 & 0.06 & 0.3394 \\
 & surf & 109.65 & 107.03 & 2.4$^{b}$ & 0.45 & 0.07 & 0.6457 \\
 & mix & 3.5873 & 3.6774 & 3.0 & 5.2 & 0.44 & 0.1223 \\
$1g_{9/2}$ & vol & 0.13464 & 0.13296 & 1.3$^{b}$ & 0.84 & 0.03 & 0.6299 \\
 & surf & $1.5886\times10^{-5}$ & $4.7328\times10^{-5}$ & $\times2.98$$^{a}$ & 39.6$^{a}$ & 0.11 & 0.0175 \\
 & mix & 0.069227 & 0.068566 & 0.96$^{b}$ & 0.81 & 0.07 & 0.4907 \\
$2p_{3/2}$ & vol & 8.5982 & 8.5682 & 2.2 & 3.6 & 0.10 & 0.2239 \\
 & surf & 109.48 & 107.10 & 2.2$^{b}$ & 0.34 & 0.06 & 0.7531 \\
 & mix & $7.8736\times10^{-4}$ & $2.4082\times10^{-3}$ & $\times3.91$$^{a}$ & $\times14.7$$^{a}$ & 0.15 & 0.0026 \\
$1f_{5/2}$ & vol & 1.6931 & 1.6863 & 2.9 & 7.1 & 0.10 & 0.1077 \\
 & surf & 118.28 & 116.67 & 1.4$^{b}$ & 0.18 & 0.05 & 0.8201 \\
 & mix & 1.4102 & 1.3791 & 2.9 & 6.3 & 0.05 & 0.0958 \\
$1f_{7/2}$ & vol & 0.39722 & 0.40353 & 5.3 & 13.6 & 0.07 & 0.0578 \\
 & surf & 105.95 & 104.87 & 1.0$^{b}$ & 0.07 & 0.04 & 0.8439 \\
 & mix & 6.7949 & 6.7546 & 0.98 & 2.5 & 0.01 & 0.2327 \\
$2s_{1/2}^{(2)}$ & vol & 1.0216 & 1.1021 & 8.7 & 9.2 & 0.11 & 0.0677 \\
 & surf & 19.915 & 20.570 & 3.3$^{b}$ & 0.65 & 1.9 & 0.4079 \\
 & mix & 3.5945 & 3.7156 & 3.4$^{b}$ & 2.8 & 0.08 & 0.1357 \\
$1d_{3/2}^{(2)}$ & vol & 2.0292 & 2.1297 & 5.0 & 5.9 & 0.10 & 0.1053 \\
 & surf & 8.3536 & 8.8867 & 6.4$^{b}$ & 0.92 & 0.40 & 0.3006 \\
 & mix & 3.5307 & 3.6656 & 3.8$^{b}$ & 2.6 & 0.14 & 0.1473 \\
$1d_{5/2}^{(2)}$ & vol & 1.4780 & 1.5591 & 5.5$^{b}$ & 5.3 & 0.12 & 0.0959 \\
 & surf & 0.32318 & 0.45106 & 39.6$^{b}$ & 4.8 & 0.15 & 0.0600 \\
 & mix & 1.5054 & 1.5975 & 6.1$^{b}$ & 2.7 & 0.35 & 0.1013 \\
\end{tabular}
\end{ruledtabular}
\end{table*}

In Table~\ref{tab:reg} and Fig.~\ref{fig:reg}, we compare the two schemes for the $^{120}$Sn $2p_{1/2}$ hole at the same mean gap. In the
cutoff scheme the mixed-pairing width grows monotonically with the cutoff, by a factor of $6.38$ between $E_{\rm cut}=40$ and $80$~MeV, although the mean gap is held fixed. The surface-pairing width, whose matrix element shows no cancellation ($C=0.83$), grows by $31\%$ over the same range, and the volume-pairing width decreases by $6.5\%$. Refitting the strength to $\Delta^{(3)}$ instead does not remove
this dependence, and the shape ratios inherit it (Table~\ref{tab:reg}). Varying $d$ and $2j_{\max}$ (Sec.~\ref{sec:setup}) changes the mixed
width by at most $3.6\%$, so the energy cutoff is the controlling parameter. The total energies, in contrast, change by
only $21.3$ and $89.8$~keV for volume and mixed pairing between $E_{\rm cut}=40$ and $80$~MeV ($443.2$~keV for surface pairing), within
the few hundred keV found for renormalized volume and mixed pairing in Ref.~\cite{BDNS06}. The renormalization that keeps energies and gaps
stable~\cite{BDNS06} is therefore not sufficient for a quantity in which the radial form of $\Delta(r)$ enters an amplitude. A single
value of such a ratio obtained in this scheme is meaningful only together with the cutoff, and not at all when one of the widths lies
near a zero of $M$.

With regularized pairing the dependence on the cutoff is much weaker. Between $E_{\max}=60$ and $120$~MeV the volume and mixed
widths and the two ratios vary by less than their box spreads (Table~\ref{tab:regall}; $2.7\%$ and $5.7\%$ for the ratios), so that
no cutoff dependence is resolved. In contrast, the surface width, whose box spread is only $0.45\%$, decreases monotonically by
$2.4\%$. However, this decrease stops above $120$~MeV. Between $E_{\max}=120$ and $160$~MeV the surface width varies by $0.59\%$
(from unrounded widths), whereas the volume width fluctuates by $1.3\%$.
The obtained results indicate that the surface-pairing width converges with $E_{\max}$, but more slowly than the volume one. Its value
at $E_{\max}=60$~MeV lies $2.6\%$ above the mean of the four points between $120$ and $160$~MeV.

The other $^{120}$Sn holes (Table~\ref{tab:regall}) show the same pattern for surface pairing. Between $E_{\max}=60$ and
$120$~MeV the $2p_{3/2}$, $1f_{5/2}$, $1f_{7/2}$, $2s_{1/2}$, and $1d_{3/2}$ widths change by $1.0\%$ to $6.4\%$ (three decrease and two
increase), in all five cases beyond the box spreads, and between $120$ and $160$~MeV they vary by at most $1.5\%$, less than the
corresponding volume widths ($2.2\%$ to $8.0\%$). The $1d_{5/2}$ width, in contrast, varies by $8.2\%$ even above $120$~MeV. With
$C=0.060$ this width is ill-conditioned, whereas the other five have $C\ge0.30$. The neutron Fermi energy of the surface-pairing field varies by
only $2.6$~keV between $E_{\max}=60$ and $160$~MeV. The total energy of the surface-pairing solution decreases by $146$~keV between $60$ and $120$~MeV
($20$~keV for volume pairing) and by $187$~keV up to $160$~MeV, in steps that become smaller ($1.8$~keV between $150$ and $160$~MeV).
This convergence is obtained with the rearrangement term of Eq.~(\ref{eq:rearr}). Without the factor
$(1-\alpha_qG)^{-2}$, the Fermi energy of the surface-pairing field rises from $-7.825$ to $-7.674$~MeV
between $E_{\max}=60$ and $120$~MeV, and the $1f_{5/2}$, $1f_{7/2}$, and $1d_{3/2}$ surface widths grow by up to $10.4\%$
over this range and continue to grow at $140$~MeV. In Skyrme--HFB calculations
regularization had been tested for volume and mixed pairing~\cite{BDNS06}. In relativistic Hartree--Bogoliubov calculations of the Sn
isotopes a density-dependent contact coupling, which gives a pairing field peaked at the surface, was regularized for ground-state
pairing properties~\cite{NRV05}. For mixed pairing four widths ($1g_{9/2}$, $2s_{1/2}$, $1d_{3/2}$, $1d_{5/2}$) change
slightly beyond their box spreads between $60$ and $120$~MeV, and most volume widths are stable. Except for the $2s_{1/2}$ surface width, the regularized widths, those of the $2p_{1/2}$ hole included, are more sensitive to the box than the cutoff ones. This sensitivity is governed by the cancellation index (Table~\ref{tab:regall}). Among the $22$ cells away from a zero of $M$ the box spread $\delta^{\rm reg}_{\rm box}$ varies by a factor of more than $200$, but the product $C\,\delta^{\rm reg}_{\rm box}$ by a factor of about $15$ only (from unrounded spreads).
Unlike the Fermi function of the cutoff scheme, the sharp cut of the regularized sum at $\bar e_k=E_{\max}$
(Sec.~\ref{sec:form-fields}) lets continuum states of the box enter or leave the pairing space abruptly when $R_{\rm box}$ or $E_{\max}$
is changed. In Ref.~\cite{BD05} individual continuum states entering the cutoff window are identified as the origin of the fluctuations
of the total energy with the box size, and in Ref.~\cite{BDNS06} the fluctuations of the regularized total energy with the cutoff
coincide with $(2j+1)$-fold multiplets entering the pairing window. In the widths the resulting variations of $\Delta_n(r)$ are
amplified by $1/C$ (Sec.~\ref{sec:form-gr}). This mechanism has not been isolated in the present calculations.
Widths very close to a zero of the matrix element are ill-conditioned in either scheme, as the mixed $2p_{3/2}$ width shows
(Table~\ref{tab:regall}).

At the same cutoff and mean gap, regularization multiplies $\Gamma_{\rm vol}$ by $1.70$, $\Gamma_{\rm surf}$ by $0.33$, and $\Gamma_{\rm mix}$ by $4.34$. The three different factors indicate that the two schemes differ by the radial shape of $\Delta$ and not by its
overall scale. For density-dependent pairing this is expected, since renormalized and regularized effective strengths differ by a
position-dependent factor~\cite{BDNS06}. For volume pairing that factor is close to unity, but the
self-consistent field changes as well, and the regularized $\Delta_n(r)$ is less concentrated at the surface ($Q_2=0.653$ against
$0.775$, with $Q_2$ defined in Sec.~\ref{sec:res-zero}). For mixed pairing the matrix element even changes sign,
$M_{\rm mix}=+0.0603$ (cutoff) and $-0.1293$~MeV\,fm$^{1/2}$ (regularized). The origin of the strong cutoff dependence is a zero of
$M$, which is discussed in Sec.~\ref{sec:res-zero}. However, large shape ratios are not an artifact of the cutoff. In the regularized
scheme at $E_{\max}=60$~MeV,
$\Gamma_{\rm surf}/\Gamma_{\rm vol}=$ $267$ and $69.9$ for the $1f_{7/2}$ and $1f_{5/2}$ holes.

\begin{table}[t]
\caption{The $2p_{1/2}$ hole of $^{116}$Sn and $^{124}$Sn (SLy4) in the cutoff scheme ($E_{\rm cut}=60$~MeV, calibration strengths) and in the regularized scheme ($E_{\max}=60$~MeV, default regularized strengths): mean gap $\langle\Delta_n\rangle$ (MeV), pole width $\Gamma$ (keV), and cancellation index $C$. $\delta_E'$: change (\%) of the regularized width from $E_{\max}=60$ to $120$~MeV at fixed strength (a decrease except for the mixed width of $^{116}$Sn). $\delta_{\rm box}$: spread max/min$\,-1$ (\%) of the regularized width over boxes of $25$, $30$, and $35$~fm, with the strength refitted in each box to the $30$~fm mean gap. $^{a}$Next to a zero of $M$.\label{tab:iso}}
\footnotesize\setlength{\tabcolsep}{1.3pt}
\begin{ruledtabular}
\begin{tabular}{llcccccccc}
 & & \multicolumn{3}{c}{cutoff} & \multicolumn{5}{c}{regularized} \\
 & & $\langle\Delta_n\rangle$ & $\Gamma$ & $C$ & $\langle\Delta_n\rangle$ & $\Gamma$ & $C$ & $\delta_E'$ & $\delta_{\rm box}$ \\
\hline
$^{116}$Sn & vol  & 1.3513 & 49.56 & 0.3575 & 1.3510 & 73.96 & 0.4388 & 1.17 & 1.83 \\
           & surf & 1.3122 & 430.6 & 0.7937 & 1.3240 & 120.1 & 0.5727 & 4.56 & 0.59 \\
           & mix  & 1.3332 & 0.5974 & 0.0397 & 1.3343 & 18.50 & 0.2258 & 0.13 & 3.00 \\
$^{124}$Sn & vol  & 1.1875 & 2.900 & 0.1304 & 1.2065 & 7.697 & 0.2118 & 3.23 & 3.50 \\
           & surf & 1.1556 & 246.2 & 0.8604 & 1.1908 & 103.9 & 0.7152 & 5.29 & 0.35 \\
           & mix  & 1.1705 & 5.270 & 0.1658 & 1.1917 & 0.005717 & 0.0050 & 26.3 & 125$^{a}$ \\
\end{tabular}
\end{ruledtabular}
\end{table}

For $^{116}$Sn and $^{124}$Sn (Table~\ref{tab:iso}) we use the regularized strengths of the code without refitting. At
$E_{\max}=60$~MeV the mean gaps then lie within $3.1\%$ of those of the cutoff calculations at $E_{\rm cut}=60$~MeV. Between
$E_{\max}=60$ and $120$~MeV only the surface widths change by more than their box spreads. At fixed strength, however, the surface
mean gap also decreases over this range, by $1.4\%$ in $^{116}$Sn and $1.7\%$ in $^{124}$Sn. With $\Gamma\propto\langle\Delta_n\rangle^2$,
at a fixed mean gap the surface widths would decrease by only about $2\%$, close to the $2.4\%$ found in $^{120}$Sn. As in
$^{120}$Sn, regularization increases $\Gamma_{\rm vol}$ and decreases $\Gamma_{\rm surf}$, by factors much larger than the box spreads
and the differences in mean gap, so the scheme dependence is resolved in all three isotopes. For mixed pairing each factor involves
a width next to a zero of $M$. The zero of $M$ moves from $^{116}$Sn in the cutoff scheme to $^{124}$Sn in the regularized one, where
the mixed width is ill-conditioned.

\subsection{The zero of the golden-rule matrix element}
\label{sec:res-zero}

\begin{figure}[t]
\includegraphics[width=0.9\columnwidth]{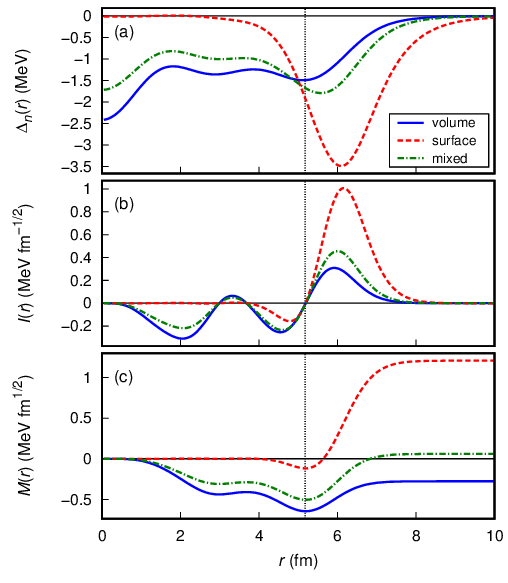}
\caption{\capSn}
\label{fig:sn}
\end{figure}

In Fig.~\ref{fig:sn}, we show the golden-rule integrand of the $2p_{1/2}$ hole for the three pairing shapes (cutoff scheme,
$E_{\rm cut}=60$~MeV, fixed mean gap). $I(r)$ changes sign four times, and the sign change at $5.17$~fm separates an interior lobe from a surface lobe of
opposite sign. Volume pairing weights the interior and ends at a negative $M$. Surface pairing, nearly zero in the
interior, ends at a positive $M$. For mixed pairing the interior contribution ($-0.505$) is almost cancelled by the surface lobe, leaving
$M=+0.060$ and $C=0.054$. The origin of the large ratio is found by exchanging wave functions and pairing fields among the three
calculations. With the wave functions of any one field, exchanging $\Delta_n(r)$ gives a golden-rule surface-to-mixed ratio between $380.1$ and
$403.6$, whereas exchanging the wave functions changes the mixed-pairing golden-rule width by at most $6\%$. The large ratio is therefore a property of $\Delta_n(r)$ and not of the wave functions. Since $M$ has opposite signs for volume and surface pairing, it must vanish for an intermediate shape.

\begin{figure*}[t]
\includegraphics[width=\textwidth]{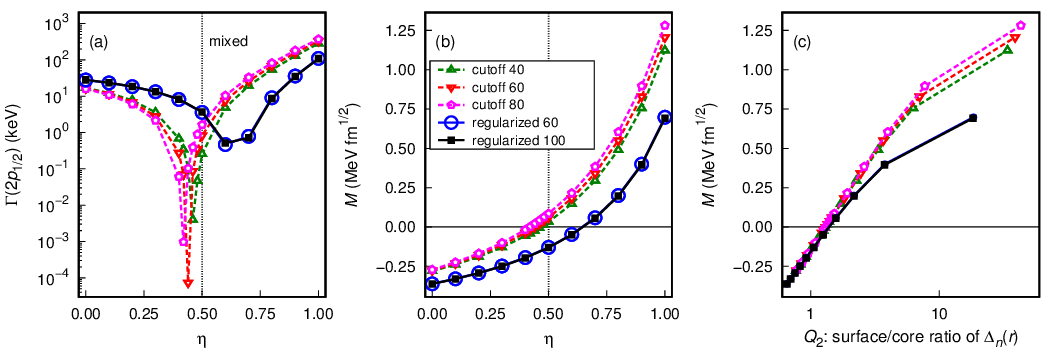}
\caption{\capZero}
\label{fig:zero}
\end{figure*}

To locate the cancellation we vary $\eta$ in Eq.~(\ref{eq:pair}) from $0$ to $1$ at fixed $\langle\Delta_n\rangle=1.2448$~MeV, in the
cutoff scheme at $E_{\rm cut}=40$, $60$, and $80$~MeV and in the regularized scheme at $E_{\max}=60$ and $100$~MeV. In every family
$M(2p_{1/2})$ changes sign once on the sampled grid (Fig.~\ref{fig:zero}, with a grid of $0.1$ refined to $0.02$ near the zero in the
cutoff families). By linear interpolation of $M$ between the two bracketing grid points the zero lies at $\eta^*=0.4652$, $0.4404$, and $0.4182$
in the cutoff scheme, i.e., it drifts by $-0.0012$ per MeV of $E_{\rm cut}$, and at $0.6447$ and $0.6470$ in the regularized scheme,
where it is stable. Near $\eta^*$ the width drops by orders of magnitude. In each cutoff family the sampled point closest to the zero lies more than
three orders of magnitude below the volume-pairing width, and the true minimum lies lower. In the regularized families the coarser grid does not resolve its depth. Away from the zero the width follows the golden rule
with a shape-dependent $M$, and $\Gamma/M^2$ is constant within $4.7\%$ wherever $C>0.1$.

The cutoff dependence of Sec.~\ref{sec:res-reg} is explained by this drift. Mixed pairing ($\eta=0.5$) lies $0.035$, $0.060$, and $0.082$ above $\eta^*$ at $40$, $60$, and $80$~MeV. With $\Gamma\propto(\eta-\eta^*)^2$ near the zero, the drift alone changes $\Gamma_{\rm mix}$ by
about $(0.082/0.035)^2=5.5$, close to the computed $6.4$. In the regularized scheme mixed pairing lies $0.14$ below a stable zero, and
$\Gamma_{\rm mix}$ is stable.

The parameter $\eta$ is only a label for the radial shape of $\Delta_n(r)$ that it produces. A simple measure of this shape is the ratio
$Q_2$ of the volume-averaged $\Delta_n$ on a surface shell $[R-1,R+1]$~fm to that on the core $[0,R-2]$~fm, with
$R=1.2A^{1/3}=5.919$~fm. At the zeros of the three cutoff families, $Q_2=1.3076$, $1.3076$, and $1.3053$. Expressed in terms of
$Q_2$, the zero does not move with $E_{\rm cut}$ (the centroid $Q_1$ defined below still drifts by $0.013$~fm), and the drift of
$\eta^*$ merely reflects the change of the shape that a given $\eta$ produces. The regularized zeros lie at somewhat larger $Q_2$
($1.394$ and $1.399$), and their centroids $Q_1=\int r^3\Delta_n\,dr/\int r^2\Delta_n\,dr$ differ from the cutoff ones ($5.463$ against
$5.547$~fm). At the zero the regularized $\Delta_n(r)$ is lower in the core and in the far tail and higher at the surface maximum.
Since $u$ and $w_h$ also change self-consistently with $\eta$, a single scalar measure of the shape cannot be expected to fix the zero
exactly.

Each channel has its own zero. In the cutoff scheme at $60$~MeV the $1f_{5/2}$ hole has its zero near $\eta=0.1$, so that volume
pairing lies close to it (the sampled width at $\eta=0.1$ is three orders of magnitude smaller than at $\eta=0$). With regularization the
zero moves to $\eta=0.317$. The deep $1d_{3/2}$ hole has a zero near $\eta=0.9$. Whether a given width is close to such a zero cannot be
told from the shape of $\Delta(r)$ alone, but it is indicated directly by the cancellation index $C$ of Eq.~(\ref{eq:C}).

\subsection{Other hole channels and other nuclei}
\label{sec:res-channels}

\begin{table*}[t]
\caption{Hole-type poles with $E_r>0$ in the search window ($0<E_h\le30$~MeV), cutoff scheme at $E_{\rm cut}=60$~MeV: mean gaps (MeV) of the three fields, resonance energy $E_r$ (MeV) of the mixed-pairing pole [Eq.~(\ref{eq:ErG})], pole widths (keV), ratio of the largest to the smallest width, and cancellation index $C$ of each shape. For $^{120,{}116,{}124}$Sn (SLy4) the calibrated strengths are used, and for $^{70}$Zn (SkM*) the volume and surface strengths are refitted to the mean gap of the mixed field. F marks channels where first-order perturbation theory fails, i.e., the golden rule deviates from the pole width by more than $10\%$ for at least one shape at full strength, so that $C$ is not used there. F refers only to the perturbative estimate. The pole widths in these rows are computed in the same way as in the other rows and do not rely on perturbation theory. The superscript $(2)$ marks the second pole of a partial wave in order of increasing $E_h$, the first one lying
below the threshold. The $1g_{9/2}$ hole of $^{116}$Sn lies below the threshold, and the $1d_{5/2}$ hole of $^{124}$Sn above the search
window ($E_h>30$~MeV), for all three shapes.\label{tab:channels}}
\begin{ruledtabular}
\begin{tabular}{llrrrrrcccc}
nucleus & hole & $E_r$ & $\Gamma_{\rm vol}$ & $\Gamma_{\rm surf}$ & $\Gamma_{\rm mix}$ & $\Gamma_{\max}/\Gamma_{\min}$ & $C_{\rm vol}$ & $C_{\rm surf}$ & $C_{\rm mix}$ & \\
\hline
$^{120}$Sn & $\langle\Delta_n\rangle$ & & 1.2455 & 1.2442 & 1.2448 & & & & & \\
 & $1d_{5/2}^{(2)}$ & 21.65 & $1.309$ & $11.08$ & $0.3382$ & 32.7 & 0.0913 & 0.3254 & 0.0491 &  \\
 & $1d_{3/2}^{(2)}$ & 18.76 & $2.215$ & $1.438$ & $2.088$ & 1.54 & 0.1118 & 0.1214 & 0.1171 &  \\
 & $2s_{1/2}^{(2)}$ & 17.22 & $1.468$ & $0.09836$ & $2.533$ & 25.8 & 0.0826 & 0.0238 & 0.1169 & F \\
 & $1f_{7/2}$ & 11.50 & $1.552$ & $100.6$ & $13.23$ & 64.8 & 0.1131 & 0.8144 & 0.3202 &  \\
 & $1f_{5/2}$ & 6.55 & $0.2137$ & $176.6$ & $8.241$ & 826.6 & 0.0380 & 0.8996 & 0.2254 &  \\
 & $2p_{3/2}$ & 5.83 & $3.628$ & $249.5$ & $4.825$ & 68.8 & 0.1442 & 0.8742 & 0.1583 &  \\
 & $2p_{1/2}$ & 4.03 & $16.46$ & $327.2$ & $0.8259$ & 396.2 & 0.2579 & 0.8319 & 0.0541 &  \\
 & $1g_{9/2}$ & 1.25 & $0.1028$ & $0.03727$ & $0.02932$ & 3.51 & 0.5663 & 0.4442 & 0.3254 & F \\
\hline $^{70}$Zn & $\langle\Delta_n\rangle$ & & 0.9554 & 0.9548 & 0.9552 & & & & & \\
 & $1p_{3/2}^{(2)}$ & 20.43 & $4.469$ & $0.9479$ & $2.843$ & 4.71 & 0.2189 & 0.1354 & 0.1925 &  \\
 & $1p_{1/2}^{(2)}$ & 17.66 & $6.369$ & $0.3988$ & $6.067$ & 16.0 & 0.2458 & 0.0903 & 0.2679 &  \\
 & $1d_{5/2}$ & 10.61 & $1.658$ & $67.31$ & $10.48$ & 40.6 & 0.1500 & 0.8822 & 0.3731 &  \\
 & $2s_{1/2}$ & 6.07 & $7.335\times10^{-3}$ & $100.9$ & $6.907$ & $1.38\times10^4$ & 0.0102 & 0.8813 & 0.3049 &  \\
 & $1d_{3/2}$ & 5.23 & $0.2689$ & $92.92$ & $4.340$ & 346 & 0.0572 & 0.9128 & 0.2241 &  \\
 & $1f_{7/2}$ & 0.55 & $0.06751$ & $1.965\times10^{-3}$ & $0.01997$ & 34.4 & 0.6472 & 0.2629 & 0.4409 & F \\
\hline $^{116}$Sn & $\langle\Delta_n\rangle$ & & 1.3513 & 1.3122 & 1.3332 & & & & & \\
 & $1d_{5/2}^{(2)}$ & 20.28 & $0.8221$ & $7.386$ & $0.4938$ & 15.0 & 0.0651 & 0.2442 & 0.0539 &  \\
 & $1d_{3/2}$ & 17.33 & $1.162$ & $0.01434$ & $2.417$ & 168 & 0.0729 & 0.0082 & 0.1147 & F \\
 & $2s_{1/2}^{(2)}$ & 16.04 & $9.903\times10^{-3}$ & $0.7380$ & $1.446$ & 146 & 0.0064 & 0.0737 & 0.0788 & F \\
 & $1f_{7/2}$ & 10.06 & $0.1716$ & $123.9$ & $10.40$ & 722 & 0.0364 & 0.8475 & 0.2729 &  \\
 & $1f_{5/2}$ & 5.00 & $3.444$ & $233.7$ & $4.948$ & 67.8 & 0.1326 & 0.8888 & 0.1524 &  \\
 & $2p_{3/2}$ & 4.55 & $15.14$ & $305.1$ & $1.117$ & 273 & 0.2506 & 0.8464 & 0.0659 &  \\
 & $2p_{1/2}$ & 2.77 & $49.56$ & $430.6$ & $0.5974$ & 721 & 0.3575 & 0.7937 & 0.0397 &  \\
\hline $^{124}$Sn & $\langle\Delta_n\rangle$ & & 1.1875 & 1.1556 & 1.1705 & & & & & \\
 & $1d_{3/2}^{(2)}$ & 20.34 & $5.087$ & $4.277$ & $2.554$ & 1.99 & 0.1853 & 0.2356 & 0.1436 &  \\
 & $2s_{1/2}^{(2)}$ & 18.45 & $5.633$ & $2.026$ & $3.972$ & 2.78 & 0.1836 & 0.1333 & 0.1660 &  \\
 & $1f_{7/2}$ & 12.93 & $4.723$ & $74.87$ & $16.53$ & 15.9 & 0.2007 & 0.7666 & 0.3729 &  \\
 & $1f_{5/2}$ & 8.16 & $0.5541$ & $129.9$ & $11.90$ & 234 & 0.0659 & 0.9026 & 0.3007 &  \\
 & $2p_{3/2}$ & 7.11 & $0.02744$ & $194.2$ & $10.08$ & $7.08\times10^3$ & 0.0146 & 0.8870 & 0.2626 &  \\
 & $2p_{1/2}$ & 5.35 & $2.900$ & $246.2$ & $5.270$ & 84.9 & 0.1304 & 0.8604 & 0.1658 &  \\
 & $1g_{9/2}$ & 2.67 & $2.069$ & $3.960$ & $0.5036$ & 7.86 & 0.4981 & 0.5587 & 0.2364 & F \\
\end{tabular}
\end{ruledtabular}
\end{table*}

\begin{figure}[t]
\includegraphics[width=0.9\columnwidth]{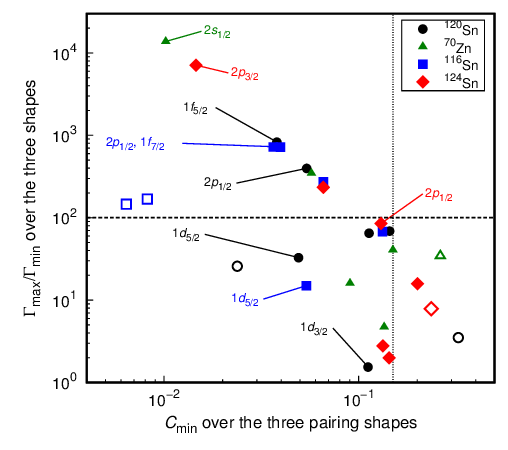}
\caption{\capChan}
\label{fig:chan}
\end{figure}

Table~\ref{tab:channels} and Fig.~\ref{fig:chan} extend the analysis to all $28$ hole-type poles with $E_r>0$ in the search window of $^{120}$Sn,
$^{116}$Sn, $^{124}$Sn (SLy4), and $^{70}$Zn (SkM*). In six channels, marked F in Table~\ref{tab:channels}, first-order perturbation theory is not adequate at full pairing
strength, and $C$ is not used there. In these channels the golden rule still reproduces $\Gamma/s^2$ at $s=0.05$ (Sec.~\ref{sec:setup}), so the deviation at full strength reflects higher orders in $\Delta$. The pole widths of these channels are not affected. The six channels are of two kinds. Three are high-$l$ holes close to the threshold ($1g_{9/2}$ of $^{120}$Sn and $^{124}$Sn and $1f_{7/2}$ of $^{70}$Zn), whose widths are very sensitive to the energy shift induced by the pairing field. The other three ($2s_{1/2}$ of $^{120}$Sn, $1d_{3/2}$ and $2s_{1/2}$ of $^{116}$Sn) have $C_{\min}<0.025$, so that the first-order amplitude itself is strongly suppressed. Among the $22$ remaining channels large shape ratios are common. In fifteen of them $M$ has opposite signs for volume and surface pairing, so that a zero of $M$ lies between $\eta=0$ and $1$. Nine
have $\Gamma_{\max}/\Gamma_{\min}>100$, and the two largest ratios ($^{70}$Zn $2s_{1/2}$ and $^{124}$Sn $2p_{3/2}$) both arise from a
small volume-pairing width with $C<0.015$. The $^{120}$Sn $2p_{1/2}$ hole is therefore
not an exceptional case. The ratios do not have a definite direction either, since among these $22$ channels $\Gamma_{\rm surf}/\Gamma_{\rm vol}$ ranges from $0.0626$ ($^{70}$Zn $1p_{1/2}$) to $1.38\times10^4$.

The largest $C_{\min}$ among the nine channels with a ratio above $100$ is $0.0659$ ($^{116}$Sn $2p_{3/2}$ and
$^{124}$Sn $1f_{5/2}$). Note that, since
$\Gamma\propto C^2(\int|I|\,dr)^2$, a ratio above $100$ requires a ratio of $C$ above about $10$ unless $\int|I|\,dr$ changes strongly
between the shapes, so that with $C\le1$ the minimum $C$ is almost necessarily small. The extreme sensitivities therefore
originate in the matrix element $M$ and not in the size of the integrand. A small $C_{\min}$ is necessary but not sufficient. The
$1d_{5/2}$ holes of $^{120}$Sn and $^{116}$Sn have $C_{\min}$ close to that of the $^{120}$Sn $2p_{1/2}$ hole but much smaller ratios. Together with the
$\eta$ families of Sec.~\ref{sec:res-zero}, this identifies the mechanism as the vicinity of a zero of $M$ for one of the standard
shapes.

The cancellation for mixed pairing in the $2p_{1/2}$ channel is accidental. It is even stronger in $^{116}$Sn, although there the mixed-pairing golden rule deviates from the pole width by $8.8\%$, and
absent in $^{124}$Sn. The ratios of this subsection refer to the cutoff
scheme. For channels close to a zero of $M$ they depend on the scheme (Sec.~\ref{sec:res-reg}). With regularized pairing the two
isotopes exchange their roles (Table~\ref{tab:iso}).

\subsection{Comparison with Pei \textit{et al.}}
\label{sec:res-pei}

\begin{figure}[t]
\includegraphics[width=0.95\columnwidth]{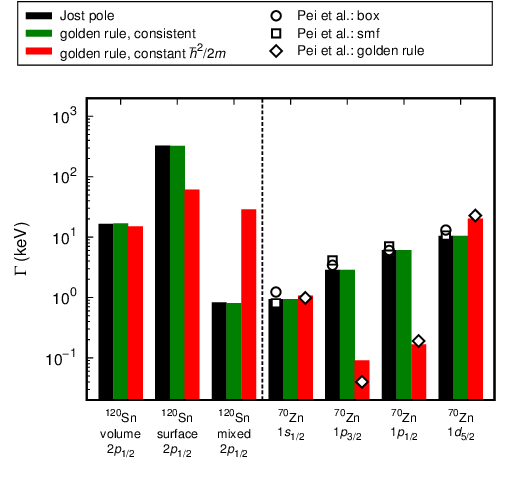}
\caption{\capPei}
\label{fig:pei}
\end{figure}

For deep neutron holes in $^{70}$Zn (SkM*, mixed pairing), Pei, Kruppa, and Nazarewicz obtained perturbative widths of $0.04$ and
$0.19$~keV for the $1p_{3/2}$ and $1p_{1/2}$ holes, against $3.42$ and $6.01$~keV
from box stabilization in their Table~I~\cite{PKN11}. For the deeper $1s_{1/2}$ hole, in contrast, their perturbative width of $0.98$~keV lies between the box-stabilized ($1.23$~keV) and smoothing--fitting ($0.80$~keV) values. Their continuum functions were computed with a constant $\hbar^2/2m=20.734$~MeV\,fm$^2$,
and they attributed the discrepancy to the sensitivity of the overlap integral to the continuum function for surface-peaked
pairing~\cite{PKN11}. Our pairing strength, the HFBRAD value for SkM* ($t_0'=-233.9$~MeV\,fm$^3$~\cite{BD05}), differs from theirs
($-234.85$~MeV\,fm$^3$) by $0.4\%$. Our quasiparticle pole energies ${\rm Re}\,E$ of the four holes agree with theirs within $8$~keV. The $1s_{1/2}$ hole lies above the search window of Table~\ref{tab:channels} and is added for this comparison. Reference~\cite{PKN11} takes the continuum function as the scattering solution of the HF
equation but does not specify how the HF potential is treated when the effective mass is replaced by the constant. In our evaluation the central and spin--orbit fields of the HF Hamiltonian are kept and only $B(r)$ is replaced by the constant, while the hole wave function is that of the full HF Hamiltonian.

Figure~\ref{fig:pei} presents the comparison. The consistent golden rule reproduces our pole widths of the four $^{70}$Zn holes within $0.3\%$. If instead the continuum function is computed with the constant $\hbar^2/2m$ of
Ref.~\cite{PKN11}, the golden rule gives $0.090$ and $0.166$~keV for the two $p$ holes, more than an order of magnitude below the pole
widths, to be compared with $0.04$ and $0.19$~keV in Table~I of Ref.~\cite{PKN11}. For the $1d_{5/2}$ hole the sign of the
deviation is reversed ($20.1$ against $10.48$~keV, and $22.7$~keV in Pei \textit{et al.}). This reproduces the direction and order of magnitude of their discrepancy. For the $1s_{1/2}$ hole the same evaluation gives $1.14$ times our pole width of $0.938$~keV, so that the perturbative expression does not fail there, as observed in Ref.~\cite{PKN11}. The constant-mass continuum function thus accounts both for the failure of the perturbative widths of the $p$ holes and for its absence for the $1s_{1/2}$ hole. For the $p$ and $d$ holes, deviations of the same direction and order of magnitude are obtained when the constant is set to the asymptotic value $B_\infty$ instead of
the value of Ref.~\cite{PKN11}. Therefore, it is the neglect of the position dependence of the effective mass, and not the value of the constant,
that matters. With a constant mass the continuum function is in general no longer orthogonal to the hole state, so that a
constant part of $\Delta_n(r)$ also contributes to $M$ (Sec.~\ref{sec:form-gr}). In this evaluation the cancellation index of the $p$ holes drops from the values of Table~\ref{tab:channels} to
$0.035$ and $0.047$. These results support the explanation of Ref.~\cite{PKN11} and show that the continuum function should be
generated by the same Hamiltonian as the hole state. Inside the nucleus, $\hbar^2/2m^*$ exceeds its asymptotic value by $28\%$ in
$^{70}$Zn and by up to $53\%$ in $^{120}$Sn. For the $^{120}$Sn $2p_{1/2}$ hole the same approximation gives between $0.186$ (surface) and $34.8$ (mixed) times the pole width (Fig.~\ref{fig:pei}), and the mixed-pairing cancellation index rises from $0.054$ to $0.451$. Our
pole widths and the box-stabilized values of Ref.~\cite{PKN11} differ by factors of $0.99$ to $1.31$ (box/pole). The box-stabilized widths
were extracted from the box-size dependence of the HFBRAD eigenvalues, the
smoothing--fitting widths from the discretized spectrum of the two-dimensional code HFB-AX, and the two differ among themselves by $16$--$54\%$~\cite{PKN11}. The comparison therefore separates the two sources of the discrepancy. The perturbative widths agree with the poles once the
continuum function is generated with the effective mass of the hole Hamiltonian, and the remaining differences between the box-stabilized values and our poles are comparable with the spread between the two box-based estimates of Ref.~\cite{PKN11}.

\section{Discussion}
\label{sec:disc}
The width computed here is the escape width of a quasiparticle on a frozen core. Measured widths of deep-hole structures in tin are of
the order of MeV. For the $1g_{9/2}$ neutron hole of $^{119}$Sn the main peak lies at $5.45$~MeV with a spreading width of
$1.45$~MeV, damping widths of $3$--$5$~MeV are quoted for hole states at $4$--$10$~MeV~\cite{GSV88}. The
$\ell=1$ hole strength in $^{115}$Sn appears in two broad structures of $1.9$ and $2.85$~MeV full width at half
maximum~\cite{vdW77}. These widths arise mainly from the coupling to more complex configurations. In heavy nuclei the escape
width is ``only a small fraction of the observed width''~\cite{GSV88}, and Dobaczewski \textit{et al.}\ judged it very unlikely
to be measured separately~\cite{DNW96}. In the regularized scheme our $^{120}$Sn $2p_{1/2}$ widths lie between $3.6$ and
$110$~keV for the three standard pairing shapes, of the same order as the estimates quoted in Sec.~\ref{sec:intro}. Therefore, the
present widths cannot be compared directly with measured ones. However, they show how escape widths, and more generally the continuum
couplings induced by pairing, should be computed.

First, an escape width should be computed in a pairing scheme in which it is stable against the cutoff. The
renormalization that stabilizes energies and gaps~\cite{BDNS06} does not achieve this, because the radial form of $\Delta(r)$ enters an
amplitude. Regularization achieves it to a large extent. For the volume and mixed $2p_{1/2}$ widths no cutoff dependence remains within the box spreads, and the residual dependence of most other channels is at the level of a few percent. The surface-pairing widths converge more slowly. For the six well-conditioned holes they change by up to $6.4\%$ between $E_{\max}=60$ and $120$~MeV and by at most $1.5\%$ between $120$ and $160$~MeV, with the rearrangement term of Eq.~(\ref{eq:rearr}). Second, the radial shape of a zero-range pairing force is a model choice that a single value of
$\Delta^{(3)}$ or of the mean gap does not constrain, and it changes escape widths by large, channel-dependent factors. The shape ratios quoted here refer to fields normalized to the same density-weighted mean gap, Eq.~(\ref{eq:gap}), and with the $\Delta^{(3)}$ anchor the surface-to-volume ratio of the $2p_{1/2}$ widths is roughly halved (Table~\ref{tab:reg}). The statement that
surface pairing gives larger widths ``in general''~\cite{DNW96} holds for six of the eight $^{120}$Sn holes in the regularized
scheme but not for every channel. Third, near a zero of $M$ the width is ill-conditioned. The pole itself is computed accurately, but
small changes of the pairing shape, of the cutoff, or of the box change it by large factors. The cancellation index $C$ is easily evaluated and could accompany any
computed escape width. In the cases studied here, for $C<0.1$ only the order of magnitude of the width is significant.

When the continuum function and the hole are generated by the same
Hamiltonian, the golden rule reproduces the pole widths of the four $^{70}$Zn holes within $0.3\%$, and at reduced pairing strength it
reproduces $\Gamma/s^2$ within $2.0\%$ in every channel examined here. A continuum function computed with a constant $\hbar^2/2m$ changes
the balance between the interior and surface lobes of $I(r)$, and the resulting error exceeds an order of magnitude for the $^{70}$Zn
$p$ holes, although their consistent integrands show no strong cancellation. For the $1s_{1/2}$ hole the error is $14\%$, in line with Table~I of Ref.~\cite{PKN11}. For the $^{70}$Zn holes, the failure of the
perturbative expression reported in Ref.~\cite{PKN11} is thus accounted for in our calculations by the continuum function, and not by
first-order perturbation theory itself. Beyond first order, in six of the $28$ channels, the pole calculation
remains necessary.

A measurement of escape widths would have to rely on channels or nuclei where the spreading is weaker than for deep holes in tin.
Shallow holes near the drip line are natural candidates. For $^{21}$C, Kobayashi and Matsuo compared quasiparticle-resonance widths
of a Woods--Saxon HFB model with preliminary peak energies and widths from neutron knockout on $^{22}$C~\cite{KM23}.

\section{Conclusions}
\label{sec:sum}
We have studied the escape widths of deep neutron holes in $^{116,{}120,{}124}$Sn and $^{70}$Zn from the Jost-function poles of the
HFB equations on self-consistent Skyrme fields, with cutoff and regularized pairing and a continuous family of pairing shapes. With
a cutoff and a strength renormalized to a fixed mean gap, the surface-pairing width of the $^{120}$Sn $2p_{1/2}$ hole grows by $31\%$
between $E_{\rm cut}=40$ and $80$~MeV, and the mixed-pairing width by a factor of $6.4$. With regularized pairing the volume and mixed
widths of this hole show no cutoff dependence within their box spreads. The well-conditioned surface-pairing
widths change by up to $6.4\%$ between $E_{\max}=60$ and $120$~MeV and by at most $1.5\%$ between $120$ and $160$~MeV. The obtained
results indicate that the sensitivity is caused by a zero of the golden-rule matrix element as a function of the radial shape of the
pairing field, near which the width drops by orders of magnitude. The cancellation index $C$ signals this situation. A golden rule
evaluated with the continuum function of the same Hamiltonian as the hole reproduces the pole widths of four deep $^{70}$Zn holes
within $0.3\%$. With a constant $\hbar^2/2m$ it reproduces both the failure of the perturbative widths of the $1p$ holes and their
validity for the $1s_{1/2}$ hole found by Pei \textit{et al.}

Although the escape width of a deep hole is far smaller than its spreading width, it is sensitive to the radial structure of the
pairing functional and to the treatment of the pairing space. Total energies and mean gaps are much less sensitive to
them~\cite{BDNS06}. The present study is restricted to spherical even--even cores and to one functional per nucleus. In the near
future, we plan to perform regularized calculations with refitted strengths for other nuclei and at larger cutoffs, and to extend the
method to shallow holes near the drip line, where escape widths may become accessible to experiment. This work is a step toward the
description of neutron elastic scattering by open-shell nuclei with the Jost--HFB method on fully self-consistent fields.

\begin{acknowledgments}
This work was supported by Hue University under the Core Research Program, Grant No. NCTB.DHH.2025.17. T. V. Nhan Hao also
acknowledges the support from Vietnam Ministry of Education and Training (MOET) under Grant No.~B2027-DHH-14.
\end{acknowledgments}

\section*{Data availability}
The data that support the findings of this article are not publicly available upon publication because it is not technically
feasible and/or the cost of preparing, depositing, and hosting the data would be prohibitive within the terms of this research
project. The data are available from the authors upon reasonable request.

\end{document}